\magnification \magstep1
\raggedbottom
\openup 2\jot
\voffset6truemm
\centerline {\bf PEIERLS BRACKETS: FROM FIELD THEORY}
\centerline {\bf TO DISSIPATIVE SYSTEMS}
\vskip 1cm
\noindent
Giuseppe Bimonte, Giampiero Esposito, Giuseppe Marmo, 
Cosimo Stornaiolo
\vskip 1cm
\noindent
{\it Dipartimento di Scienze Fisiche, Universit\`a di Napoli Federico II,
Complesso Universitario di Monte S. Angelo, Via Cintia, Edificio N',
80126 Napoli, Italy}
\vskip 0.3cm
\noindent
{\it Istituto Nazionale di Fisica Nucleare, Sezione di Napoli,
Complesso Universitario di Monte S. Angelo, Via Cintia, Edificio N',
80126 Napoli, Italy}
\vskip 1cm
\noindent
{\bf Abstract}.
Peierls brackets are part of the space-time approach to
quantum field theory, and provide a Poisson bracket which, being
defined for pairs of observables which are group invariant, is
group invariant by construction. It is therefore well suited for
combining the use of Poisson brackets and the full diffeomorphism
group in general relativity. The present paper provides at first an 
introduction to the topic, with applications to gauge field theory.
In the second part, a
set of brackets for classical dissipative systems, subject to
external random forces, are derived. The method is inspired by the
old procedure of Peierls, for deriving the canonical
brackets of conservative systems, starting from an action
principle. It is found that an adaptation of Peierls' method is
applicable also to dissipative systems, when the friction term can
be described by a linear functional of the coordinates, as is the
case in the classical Langevin equation, with an arbitrary memory
function. The general expression for the brackets satisfied by
the coordinates, as well as by the external random forces, at
different times, is determined, and it turns out that they all
satisfy the Jacobi identity. Upon quantization, these classical
brackets  are found to coincide with the commutation rules for the
quantum Langevin equation, that have been obtained in the past,
by appealing to microscopic conservative quantum models for the
friction mechanism.
\vskip 1cm
\leftline {\bf 1. Introduction}	
\vskip 0.3cm
\noindent
Although the Hamiltonian formalism provides a powerful tool for
studying general relativity [1], its initial-value problem and
the approach to canonical quantization [2], it suffers from
severe drawbacks: the space $+$ time split of $(M,g)$ disagrees
with the aims of general relativity, and the space-time topology
is taken to be $\Sigma \times {\bf R}$, so that the full diffeomorphism
group of $M$ is lost [3,4].

However, as was shown by DeWitt in the sixties [5], it remains
possible to use a Poisson-bracket formalism which preserves the
full invariance properties of the original theory, by relying upon
the work of Peierls [6]. In our paper, whose aims are pedagogical,
we begin by describing the general framework, assuming that the
reader has been introduced to the DeWitt covariant approach
to quantum field theory [5].
Let us therefore consider a gauge field theory with classical 
action functional $S$ and generators of infinitesimal gauge
transformations denoted by $R_{\; \alpha}^{i}$. The small
disturbances $\delta \varphi^{i}$ are ruled by the invertible
differential operator
$$
F_{ij} \equiv S_{,ij}+\gamma_{ik}R_{\; \alpha}^{k}
{\widetilde \gamma}^{\alpha \beta}\gamma_{jl}R_{\; \beta}^{l}, 
\eqno (1)
$$
where $\gamma_{ij}$ is a local and symmetric matrix which is
taken to transform like $S_{,ij}$ under group transformations,
and ${\widetilde \gamma}^{\alpha \beta}$ is a local, non-singular,
symmetric matrix which transforms according to the adjoint 
representation of the infinite-dimensional invariance group
(hence one gets $R_{i \alpha} \equiv \gamma_{ij}R_{\; \alpha}^{j}$
and $R_{i}^{\; \alpha} \equiv {\widetilde \gamma}^{\alpha \beta}
R_{i \beta}$, respectively). We are interested in advanced and
retarded Green functions $G^{\pm}$ which are left inverses of
$-F$, i.e.
$$
G^{\pm ij}F_{jk}=-\delta_{\; k}^{i}.
\eqno (2)
$$
Furthermore, the form of $F_{ij}$ and arbitrariness of Cauchy data
imply that $G^{\pm}$ are right inverses as well, i.e.
$$
F_{ij}G^{\pm jk}=-\delta_{i}^{\; k}.
\eqno (3)
$$
If symmetry of $F$ is required, one also finds
$$
G^{+ij}=G^{-ji}, \; G^{-ij}=G^{+ji},
\eqno (4)
$$
because in general
$$
G^{\pm ij}-G^{\mp ji}=G^{\pm ik}(F_{kl}-F_{lk})G^{\mp jl}.
\eqno (5)
$$
Thus, the {\it supercommutator function} defined as
$$
{\widetilde G}^{ij} \equiv G^{+ij}-G^{-ij}
\eqno (6)
$$
is antisymmetric in that ${\widetilde G}^{ij}=-
{\widetilde G}^{ji}$.
These properties show that, on defining
$\delta_{A}^{\pm}B \equiv \varepsilon B_{,i}G^{\pm ij}A_{,j}$,
one has, on relabelling dummy indices,
$$
\delta_{A}^{\pm}B
=\varepsilon B_{,j}G^{\pm ji}A_{,i}
=\varepsilon A_{,i}G^{\mp ij}B_{,j} 
=\delta_{B}^{\mp}A.
\eqno (7)
$$ 
These are the {\it reciprocity relations}, which express the idea 
that the retarded (resp. advanced) effect of 
$A$ on $B$ equals the advanced (resp. retarded) effect of $B$
on $A$. Another cornerstone of the formalism is a relation
involving the Green function $\widehat G$ of the operator
$-{\widehat F}$, having set $R_{k \beta}R_{\; \alpha}^{k}
\equiv {\widehat F}_{\beta \alpha}$; this is
$$
R_{\; \alpha}^{i} \; {\widehat G}^{\pm \alpha \beta} \;
{\widetilde \gamma}_{\beta \delta}
=R_{\; \alpha}^{i} \; {\widehat G}_{\; \; \; \delta}^{\pm \alpha} 
=G^{\pm ij} \; \gamma_{jk} \; R_{\; \delta}^{k}
=G^{\pm ij} \; R_{j \delta}.
\eqno (8)
$$ 
This holds because, {\it for background fields satisfying the field
equations}, one finds that
$$
F_{ik} R_{\; \alpha}^{k}=R_{i}^{\; \beta}R_{k \beta}R_{\; \alpha}^{k}
=R_{i}^{\; \beta}{\widehat F}_{\beta \alpha}.
\eqno (9)
$$
On multiplying this equation on the left by $G^{\pm ji}$ and on the
right by ${\widehat G}^{\pm \alpha \beta}$ one gets
$$
R_{\; \alpha}^{j}{\widehat G}^{\pm \alpha \beta}
=G^{\pm ji}R_{i}^{\; \beta},
\eqno (10)
$$
i.e. the desired formula (8) is proved. Moreover, by virtue of
(4), the transposed equations 
$$
{\widehat G}^{\pm \alpha \beta}R_{\; \beta}^{j}
=R_{i}^{\; \alpha}G^{\pm ij}
\eqno (11)
$$
also hold.
We are now in a position to define the Peierls bracket of any two
observables $A$ and $B$. First, we consider the operation
$$
D_{A}B \equiv \lim_{\varepsilon \to 0}\varepsilon^{-1}
\delta_{A}^{-}B,
\eqno (12)
$$
with $D_{B}A$ obtained by interchanging $A$ with $B$ in (12).
The {\it Peierls bracket} of $A$ and $B$ is then defined by
$$
(A,B) \equiv D_{A}B-D_{B}A=\lim_{\varepsilon \to 0}
{1\over \varepsilon}\Bigr[\varepsilon A_{1}G^{+}B_{1}
-\varepsilon A_{1}G^{-}B_{1}\Bigr] 
=A_{1}{\widetilde G}B_{1}=A_{,i}{\widetilde G}^{ij}B_{,j},
\eqno (13)
$$ 
where we have used (7) and (12) to obtain the last expression.
Following DeWitt [7], it should be stressed that the Peierls 
bracket depends only on the behaviour of infinitesimal disturbances.

In classical mechanics, following Peierls [6], we may arrive at the
derivatives in (12) and (13) starting from the action functional
$S \equiv \int L \; d\tau$ and considering the extremals of $S$ and
those of $S + \lambda A$, where $\lambda$ is an infinitesimal 
parameter and $A$ any function of the path $\gamma$. Next we
consider solutions of the modified equations as expansions in powers
of $\lambda$, and hence the new set of solutions to first order
reads
$$
\gamma'(\tau)=\gamma(\tau)+\lambda D_{A} \gamma(\tau).
\eqno (14)
$$
This modified solution is required to obey the condition 
that, in the distant past, it should be identical with the original 
one, i.e.
$$
D_{A}\gamma(\tau) \rightarrow 0 \; {\rm as} \; \tau 
\rightarrow -\infty.
\eqno (15)
$$
Similarly to the construction of the above ``retarded'' solution,
we may define an ``advanced'' solution
$$
\gamma''(\tau)=\gamma(\tau)+\lambda {\cal D}_{A} \gamma(\tau),
\eqno (16)
$$
such that
$$
{\cal D}_{A}\gamma(\tau) \rightarrow 0 \; {\rm as} \; 
\tau \rightarrow + \infty.
\eqno (17)
$$
From these modified solutions one can now find $D_{A}\gamma(\tau)$
along the solutions of the un-modified action and therefore,
to first order,
the changes in any other function $B$ of the field variables, and
these are denoted by $D_{A}B$ and $D_{B}A$, respectively.
\vskip 5cm
\leftline {\bf 2. Mathematical properties of Peierls brackets}
\vskip 0.3cm
\noindent
We are now aiming to prove that
$(A,B)$ satisfies all properties of a Poisson bracket. The first
two, anti-symmetry and bilinearity, are indeed obvious:
$$
(A,B)=-(B,A),
\eqno (18)
$$
$$
(A,B+C)=(A,B)+(A,C),
\eqno (19)
$$
whereas the proof of the Jacobi identity is not obvious and is
therefore presented in detail. First, by repeated application of
(13) one finds
$$ \eqalignno{
\; &  P(A,B,C) \equiv (A,(B,C))+(B,(C,A))+(C,(A,B)) \cr 
&= A_{,i}{\widetilde G}^{il}\Bigr(B_{,j}{\widetilde G}^{jk}
C_{,k}\Bigr)_{,l}+B_{,j}{\widetilde G}^{jl}
\Bigr(C_{,k}{\widetilde G}^{ki}A_{,i}\Bigr)_{,l} 
+C_{,k}{\widetilde G}^{kl}
\Bigr(A_{,i}{\widetilde G}^{ij}B_{,j}\Bigr)_{,l} \cr 
&=  A_{,il}B_{,j}C_{,k}\Bigr({\widetilde G}^{ij}
{\widetilde G}^{kl}+{\widetilde G}^{jl}{\widetilde G}^{ki}\Bigr) 
+A_{,i}B_{,jl}C_{,k}\Bigr({\widetilde G}^{jk}
{\widetilde G}^{il}+{\widetilde G}^{kl}{\widetilde G}^{ij}\Bigr) 
\cr 
&+  A_{,i}B_{,j}C_{,kl}\Bigr({\widetilde G}^{ki}
{\widetilde G}^{jl}+{\widetilde G}^{il}{\widetilde G}^{jk}\Bigr) \cr
&+ A_{,i}B_{,j}C_{,k}\Bigr({\widetilde G}^{il}
{\widetilde G}_{\; \; \; ,l}^{jk}
+{\widetilde G}^{jl}{\widetilde G}_{\; \; \; ,l}^{ki}
+{\widetilde G}^{kl}{\widetilde G}_{\; \; \; ,l}^{ij}\Bigr).
&(20)\cr}
$$
Now the antisymmetry property of $\widetilde G$, jointly with commutation
of functional derivatives: $T_{,il}=T_{,li}$ for all $T=A,B,C$, 
implies that the first three terms on the last equality in (20)
vanish. For example one finds
$$ \eqalignno{
\; & A_{,il}B_{,j}C_{,k}\Bigr({\widetilde G}^{ij}
{\widetilde G}^{kl}+{\widetilde G}^{jl}{\widetilde G}^{ki}\Bigr)
=A_{,li}B_{,j}C_{,k}\Bigr({\widetilde G}^{lj}
{\widetilde G}^{ki}+{\widetilde G}^{ji}{\widetilde G}^{kl}\Bigr) \cr
&= -A_{,il}B_{,j}C_{,k}\Bigr({\widetilde G}^{jl}
{\widetilde G}^{ki}+{\widetilde G}^{ij}{\widetilde G}^{kl}\Bigr)=0,
&(21)\cr}
$$
and an entirely analogous procedure can be applied to the terms 
containing the second functional derivatives $B_{,jl}$ and
$C_{,kl}$. The last term in (20) requires new calculations
because it contains functional derivatives of ${\widetilde G}^{ij}$.
These can be dealt with after taking infinitesimal variations of
Eq. (3), so that
$$
F \; \delta G^{\pm}=-(\delta F)G^{\pm},
\eqno (22)
$$
and hence
$$
G^{\pm}F \delta G^{\pm}=F G^{\pm} \delta G^{\pm}
=-\delta G^{\pm}=-G^{\pm}(\delta F)G^{\pm},
\eqno (23)
$$
i.e.
$$
\delta G^{\pm}=G^{\pm}(\delta F)G^{\pm}.
\eqno (24)
$$
Thus, the desired functional derivatives of advanced and retarded
Green functions read
$$ \eqalignno{
\; & G_{\; \; \; \; \; ,c}^{\pm ij}=G^{\pm ia}F_{ab,c}G^{\pm bj} 
=G^{\pm ia}\Bigr(S_{,ab}+R_{a \alpha}R_{b}^{\; \alpha}\Bigr)_{,c}
G^{\pm bj} \cr 
&= G^{\pm ia}S_{,abc}G^{\pm bj}+G^{\pm ia}R_{a \alpha,c}
R_{b}^{\; \alpha} \; G^{\pm bj} 
+G^{\pm ia}R_{a \alpha} R_{b \; \; ,c}^{\; \alpha}
\; G^{\pm bj}.
&(25)\cr}
$$
In this formula the contractions $R_{b}^{\; \alpha} \; G^{\pm bj}$
and $G^{\pm ia}R_{a \alpha}$ can be re-expressed with the help of
Eqs. (10) and (11), and eventually one gets
$$ 
G_{\; \; \; \; \; ,c}^{\pm ij}=G^{\pm ia} S_{,abc}G^{\pm bj}
+G^{\pm ia}R_{a \alpha,c}{\widehat G}^{\pm \alpha \beta}
R_{\; \beta}^{j} 
+R_{\; \beta}^{i} \; {\widehat G}_{\; \; \; \alpha}^{\pm \beta}
\; R_{b \; \; ,c}^{\; \alpha} \; G^{\pm bj}.
\eqno (26)
$$ 
By virtue of the group invariance property satisfied by all
physical observables, the second and third term on the right-hand
side of Eq. (26) give vanishing contribution to (20). One is
therefore left with the contributions involving third functional
derivatives of the action. Bearing in mind that $S_{,abc}=
S_{,acb}=S_{,bca}=...$, one can relabel indices summed over, finding
eventually (upon using (4))
$$ \eqalignno{
\; & P(A,B,C)=A_{,i}B_{,j}C_{,k}\Bigr[(G^{+ic}-G^{-ic})
(G^{+ja}G^{+bk}-G^{-ja}G^{-bk}) \cr 
&+ (G^{+jc}-G^{-jc})(G^{+ka}G^{+bi}-G^{-ka}G^{-bi}) \cr 
&+ (G^{+kc}-G^{-kc})(G^{+ia}G^{+bj}-G^{-ia}G^{-bj})\Bigr]S_{,abc} \cr
&= A_{,i}B_{,j}C_{,k}\Bigr[(G^{+ia}-G^{-ia})(G^{+jb}G^{-kc}
-G^{-jb}G^{+kc})\cr 
&+ (G^{+jb}-G^{-jb})(G^{+kc}G^{-ia}-G^{-kc}G^{+ia})\cr 
&+ (G^{+kc}-G^{-kc})(G^{+ia}G^{-jb}-G^{-ia}G^{+jb})
\Bigr]S_{,abc} =0.
&(27)\cr}
$$
This sum vanishes because it involves six pairs of triple products of
Green functions with opposite signs.
The Jacobi identity is therefore fulfilled. Moreover, the fourth
fundamental property of Poisson brackets, i.e.
$$
(A,BC)=(A,B)C+B(A,C)
\eqno (28)
$$
is also satisfied, because
$$ 
(A,BC)=A_{,i}{\widetilde G}^{ik}(BC)_{,k}
=A_{,i}{\widetilde G}^{ik}B_{,k}C+BA_{,i}{\widetilde G}^{ik}C_{,k} 
=(A,B)C+B(A,C).
\eqno (29)
$$ 
Thus, the Peierls bracket defined in (13) is indeed a Poisson
bracket of physical observables. Equation (28) can be regarded
as a compatibility condition of the Peierls bracket with the product
of physical observables.

It should be stressed that the idea of Peierls [6] was to introduce
a bracket related directly to the action principle without making
any reference to the Hamiltonian. This implies that even classical
mechanics should be considered as a ``field theory'' in a 
zero-dimensional space, having only the time dimension. This means
that one deals with an infinite-dimensional space of paths
$\gamma: {\bf R} \rightarrow Q$, therefore we are dealing with
functional derivatives and distributions even in this situation
where modern standard treatments rely upon $C^{\infty}$ manifolds
and smooth structures. Thus, the present treatment is hiding most
technicalities involving infinite-dimensional manifolds. In finite 
dimensions on a smooth manifold, any bracket satisfying (19) and
(28) is associated with first-order bidifferential operators [8,9];
in this proof it is important that the commutative and associative
product $BC$ is a local product. In any case these brackets at the
classical level could be a starting point to define a 
$*$-product in the spirit of non-commutative geometry [10] or
deformation quantization [11].
\vskip 0.3cm
\leftline {\bf 3. The most general Peierls bracket}
\vskip 0.3cm
\noindent
The Peierls bracket is a group invariant by construction, being
defined for pairs of observables which are group invariant, and
is invariant under both infinitesimal and finite changes in the
matrices $\gamma_{ij}$ and ${\widetilde \gamma}_{\alpha \beta}$.
DeWitt [5] went on to prove that, even if 
independent differential operators $P_{i}^{\; \alpha}$ and
$Q_{i \alpha}$ are introduced such that
$$ 
F_{ij} \equiv S_{,ij}+P_{i}^{\; \alpha}Q_{j \alpha}, \;
{\widehat F}_{\alpha \beta} \equiv Q_{i \alpha}R_{\; \beta}^{i}, \;
F_{\alpha}^{\; \beta} \equiv R_{\; \alpha}^{i} 
P_{i}^{\; \beta},
\eqno (30)
$$
are all non-singular, with unique advanced and retarded Green
functions, the reciprocity theorem expressed by (7) still
holds, and the resulting Peierls bracket is invariant under changes
in the $P_{i}^{\; \alpha}$ and $Q_{i \alpha}$, by virtue of
the identities
$$
Q_{i \alpha}G^{\pm ij}=G_{\; \; \alpha}^{\pm \; \; \beta}
R_{\; \beta}^{j},
\eqno (31)
$$
$$
G^{\pm ij}P_{j}^{\; \beta}=R_{\; \alpha}^{i}
{\widehat G}^{\pm \alpha \beta}.
\eqno (32)
$$
This is proved as follows. The composition of $F_{ik}$ with the
infinitesimal generators of gauge transformations yields
$$
F_{ik}R_{\; \alpha}^{k}=P_{i}^{\; \beta}F_{\beta \alpha},
\eqno (33)
$$
and hence
$$
G^{\pm ji}F_{ik}R_{\; \alpha}^{k}=-R_{\; \alpha}^{j}
=G^{\pm ji}P_{i}^{\; \gamma}F_{\gamma \alpha},
\eqno (34)
$$
which implies
$$
R_{\; \alpha}^{j}G^{\pm \alpha \beta}=-G^{\pm ji}P_{i}^{\; \gamma}
F_{\gamma \alpha}G^{\pm \alpha \beta}
=G^{\pm ji}P_{i}^{\; \beta},
\eqno (35)
$$
i.e. Eq. (32) is obtained. Similarly,
$$
R_{\; \alpha}^{i}F_{ij}=F_{\alpha}^{\; \beta}Q_{j \beta},
\eqno (36)
$$
and hence
$$
G_{\; \; \alpha}^{\pm \; \; \gamma}R_{\; \gamma}^{i}F_{ij}
=-Q_{j \alpha},
\eqno (37)
$$
which implies
$$ 
Q_{i \alpha}G^{\pm ij}=-G_{\; \; \alpha}^{\pm \; \; \gamma}
R_{\; \gamma}^{k}F_{ki}G^{\pm ij} 
=G_{\; \; \alpha}^{\pm \; \; \beta}R_{\; \beta}^{j},
\eqno (38)
$$
i.e. Eq. (31) is obtained. Now we use the first line of Eq. (7) 
for $\delta_{A}^{\pm}B$, jointly with Eq. (5), so that
$$ 
\delta_{A}^{\pm}B-\varepsilon B_{,i}G^{\mp ji}A_{,j} 
=\varepsilon B_{,i}R_{\; \gamma}^{i}G^{\pm \gamma \alpha}
Q_{l \alpha}G^{\mp jl}A_{,j} 
-\varepsilon B_{,i}P_{l}^{\; \alpha}G^{\pm ik}Q_{k \alpha}
G^{\mp jl}A_{,j}.
\eqno (39)
$$ 
Since $B$ is an observable by hypothesis, the first term on the right-hand 
side of (39) vanishes. Moreover one finds, from (32)
$$ 
G^{\pm ik}P_{l}^{\; \alpha}Q_{k \alpha}G^{\mp jl}
=G^{\pm il}R_{\; \beta}^{j}G^{\mp \beta \alpha}Q_{l \alpha}.
\eqno (40)
$$
and hence also the second term on the right-hand side of (39) vanishes
($A$ being an observable, for which $R_{\; \beta}^{j}A_{,j}=0$), yielding
eventually the reciprocity relation (7). Moreover, the invariance of the
Peierls bracket under variations of $P_{i \alpha}$ and $Q_{i}^{\; \alpha}$
holds because
$$ \eqalignno{
\; & \delta (\delta_{A}^{\pm}B)=\varepsilon B_{,i}
\delta G^{\pm ij}A_{,j}
=\varepsilon B_{,i}G^{\pm ik}(\delta F_{kl})G^{\pm lj}A_{,j} \cr 
&= \varepsilon B_{,i}G^{\pm ik}\Bigr[(\delta P_{k}^{\; \alpha})Q_{l \alpha}
+P_{k}^{\; \alpha}(\delta Q_{l \alpha})\Bigr]G^{\pm lj}A_{,j} \cr 
&= \varepsilon B_{,i}G^{\pm ik}(\delta P_{k}^{\; \alpha})Q_{l \alpha}
G^{\pm lj}A_{,j}+\varepsilon B_{,i}G^{\pm ik}P_{k}^{\; \alpha}
(\delta Q_{l \alpha})G^{\pm lj}A_{,j} \cr 
&= \varepsilon B_{,i}G^{\pm ik} (\delta P_{k}^{\; \alpha})
G_{\; \; \alpha}^{\pm \; \; \beta}R_{\; \beta}^{j}A_{,j} 
+\varepsilon B_{,i}R_{\; \gamma}^{i}G^{\pm \gamma \alpha}
(\delta Q_{l \alpha})G^{\pm lj}A_{,j}=0,
&(41)\cr}
$$
where Eqs. (31) and (32) have been exploited once more.
\vskip 0.3cm
\leftline {\bf 4. Quantum dissipative systems}
\vskip 0.3cm
\noindent
The study of quantum dissipative systems has attracted, over the
last decades, a lot of interest, in view of its broad spectrum of
applications, ranging from quantum optics through statistical
mechanics. The standard approach to deal with  quantum
dissipation, is based on the idea that the physical origin of
dissipation is the interaction of the system with a heat bath,
consisting of a large number of degrees of freedom. One considers
then some microscopic, conservative model for the heat bath (and
its interaction with the system), and tries  to recover the
macroscopic quantum behavior of the dissipative system alone, by
eliminating from the description the degrees of freedom describing
the bath. In Ref. [12], it is shown, indeed, that the most
general quantum Langevin equation, which is  one of the most
popular models for dissipation, can be obtained from a simple
microscopic model, where the heat bath is described by a set of
independent oscillators, linearly coupled to the system of
interest.

One may wonder whether it is possible to find a quantization
method for dissipative systems, which is based   {\it on the
macroscopic description of dissipation only}, and makes therefore
no use of microscopic models. As is well known, quantization of
dissipative systems is by no means straightforward, because in
general they admit neither a Lagrangian nor a Hamiltonian
formulation. Moreover, even in those special instances where such
a formulation can be given, the application of the conventional
canonical quantization methods leads to physically incorrect
results [13]. In this paper, we show that new classical
brackets can be consistently built for dissipative systems, by
generalizing the covariant definition of Poisson brackets for
Lagrangian systems, discovered long ago by Peierls [6]
(see also Refs. [5,14--16]). Our bracket is
defined on the infinite-dimensional functional space consisting of
all possible classical trajectories, that are accessible to the
system under the influence of the random force. It turns out that,
when dissipation is present, the random external force has a
non-vanishing bracket with the system coordinates, which implies
that it cannot be consistently taken to be zero. This seems to be
in agreement with the fluctuation-dissipation theorem, which
requires fluctuating forces, in the presence of dissipation.

By the correspondence principle, our classical brackets can be
eventually quantized, upon  substituting them by ($1/(i  \hbar)$
times) commutators. In this way, we recover the same expressions
for the commutators between the system coordinates and the random
forces, which were derived from the independent oscillator
microscopic model of Ref. [12].

In  what follows, we make no attempt at mathematical rigor, and
the presentation is totally heuristic. We hope to clarify
elsewhere the delicate issues involved in the consideration of
Poisson structures in infinite-dimensional functional spaces.
\vskip 0.3cm
\leftline {\bf 5. The classical brackets}
\vskip 0.3cm
\noindent
We consider a mechanical system, with coordinates
$(q^1,...,q^n)$, described by an action  functional  $S=\int dt
\;{\cal L}(q^i,\dot q ^i, t)$, where dot denotes a time
derivative. We assume that the Lagrangian is a polynomial of
second degree in the velocities $\dot q ^i$, and that its Hessian
$ {\partial ^2 {\cal L}}/{\partial \dot q ^i
\partial \dot  q^ j}$ is a constant, non-degenerate matrix
$M_{ij}$. We imagine that the system is in contact with a heat
bath, and we assume that the influence of the heat bath can be
described, effectively, by a mean force, characterized by a
bounded memory function $\mu_{ij}(t-t')$, and a random force
$F_i(t)$. The time  evolution of the system is then described by
the following equation of Langevin type: 
$$ -{\delta S \over 
\delta q^i(t)} +
\int_{-\infty}^t dt'\,
\mu_{ij}(t-t')\;\dot q ^j (t')=F_i(t).
\eqno (42)
$$
Here,
$\delta S/\delta q^i(t)$ denotes the functional derivative of the
action $S$: 
$$ 
{\delta S \over \delta q^i(t)} \equiv
-{d \over dt}\left({\partial {\cal L} \over \partial
\dot{q}^i}\right)+{\partial {\cal L} \over \partial q^i} .
\eqno (43)
$$
The original form of the Langevin equation results from the
singular limit, where $\mu_{ij}(t-t')$ approaches
$\gamma_{ij}\delta (t-t')$.

Mimicking the procedure found by Peierls, to compute the Poisson
brackets of a conservative Lagrangian system [6], one can
consider the effect, on the system evolution, of a small
disturbance, produced by an infinitesimal change $\bar{\delta} S$
in the form of the action. We consider changes of the form
$\bar{\delta} S= \epsilon A$, where $\epsilon$ is an infinitesimal
constant and $A$ is a local functional of the trajectory $q^j(t)$,
taken from a finite time interval. The small disturbance causes an
infinitesimal shift $\delta_A q^j(t)$ in the trajectory $q^j(t)$,
and it is easy to see that, to first order in $\epsilon$,
$\delta_A q^j$ satisfies the following linear integro-differential
equation:
$$ \eqalignno{
\; & (L\; \delta_A q)_i(t)\equiv -\int dt' 
\; {\delta^{2} S \over \delta q^i(t)
\delta q^j(t')} \; \delta_A q^j(t') \cr
&+\int_{-\infty}^t dt'\, \mu_{ij}(t-t')\; \delta_A \dot q ^j
(t')=\epsilon {\delta A \over \delta q^i(t)}, 
&(44)\cr}
$$
where it is understood that all functional derivatives are
evaluated along the undisturbed trajectory. When writing the
above equation, we have also assumed that the random force does
not undergo any variation, to first order in $\epsilon$. We point
out that, by virtue of our assumptions on the Lagrangian, the
coefficients of Eq. (44) depend only on the coordinates
$q^i(t)$ and the velocities $\dot q ^i(t)$ of the undisturbed
trajectory, while they are independent of the accelerations. This
is reassuring, because, by virtue of the random external
force, the classical trajectories possess, in general, smooth
velocities, while the acceleration does not exist, in the ordinary
sense of time-derivatives of the velocity [17].

Since the disturbance $A$ is localized in a finite time interval,
it makes sense to consider the solution $\delta ^- _A q^j(t)$ of
Eq. (44) satisfying {\it retarded} boundary conditions, i.e. 
$$
\delta ^- _A q^j(t)=0 \;\;\;\;{\rm at \;early \;times}.
\eqno (45)
$$
The non-degeneracy condition for the Hessian $M_{ij}$ of the
Lagrangian, ensures that $\delta ^- _A q^j(t)$ exists and is
unique. We consider also the {\it advanced} solution $\delta^{+}_{A}
q^j(t)$: 
$$
\delta^{+}_{A} q^j(t)=0\;\;\;\;\;{\rm at\; late\;
times}
\eqno (46)
$$ 
of the {\it adjoint} equation of Eq. (44):
$$ \eqalignno{
\; & (L^T\; \delta_A^+ q)_i(t)=-\int dt' \;  {\delta^{2} S
\over \delta q^j(t') \delta q^i(t)} \delta^{+}_{A} q^j(t') \cr
&- \int_{t}^{\infty} dt'\,\mu_{ji}(t'-t)\,\delta^+_A \dot q ^j (t')
\; =\epsilon {\delta A \over \delta q^i(t)}, 
&(47)\cr}
$$
where the superscript $T$ stands for transpose
(the transpose coincides with the adjoint, because we are in the
real field). If $B$ is another functional of the trajectory, with
support in a finite time interval, we define the bracket $\{A,B\}$
as the following expression, involving the quantities
$\delta^{\pm}_A q^j(t)$: 
$$ 
\{A,B\} \equiv {1 \over \epsilon}\int dt
{\delta B \over \delta q^i(t)} (\delta^+_A q^i(t)-\delta^-_A
q^i(t)).
\eqno (48)
$$
It is immediate to verify  that the
bracket is bilinear and satisfies the Leibniz rule:
$$
\{AB,C\}=\{A,C\}\,B+A\{B,C\},
\eqno (49)
$$
$$
\{A,BC\}=\{A,B\}\,C+B\{A,C\}.
\eqno (50)
$$
To verify that the bracket (48) is also antisymmetric and that
it satisfies the Jacobi identity, it is useful to re-express it in
terms of the Green functions $G^{\pm ij}(t,t';q)$, defined so that
$$
\delta^{\pm}_A q^i(t)= \epsilon \int dt' \, G^{\pm
ij}(t,t';q) {\delta A \over \delta q^j(t')}.
\eqno (51)
$$
The Green functions $G^{\pm ij}(t,t')$ satisfy the following boundary
conditions: 
$$
G^{-ij}(t,t';q)=0 \;\;,\;\;\;{\rm for} \;\;t \le
t',
\eqno (52)
$$
$$
G^{+ij}(t,t';q)=0 \;\;,\;\;\;\;{\rm
for} \;\;t \ge t',
\eqno (53)
$$
$$
\lim_{t \rightarrow t'^{\mp}} {\partial G^{\pm ij} \over
\partial t}(t,t';q) =\mp (M^{-1})^{ij}.
\eqno (54)
$$
We define now the {\rm commutator function}
$\widetilde{G}^{ij}(t,t';q)$: 
$$
\widetilde{G}^{ij}(t,t';q) \equiv G^{+ij}
(t,t';q)-G^{-ij}(t,t';q).
\eqno (55)
$$ 
Note that, by virtue of the
boundary conditions satisfied by the retarded and the advanced
Green functions, $\widetilde{G}^{ij}(t,t')$ and $\partial _t
\widetilde{G}^{ij}(t,t')$ are continuous, in the coincidence time
limit, $t \rightarrow t'$. By using $\widetilde{G}^{ij}(t,t')$, we
can rewrite the bracket (48) as (cf. (13)) 
$$
\{A,B\}=
\int dt  \int dt'\, {\delta B \over \delta
q^i(t)}\,\widetilde{G}^{ij}(t,t';q)\, {\delta A \over \delta
q^j(t')} .
\eqno (56)
$$
The antisymmetry of the bracket
(48) follows from the fact that the commutator
function $\widetilde{G}^{ij}$ is antisymmetric, as a consequence
of the following reciprocity relation, satisfied by the advanced
and retarded Green functions: 
$$
G^{+ ij}(t,t';q)=G^{-ji}
(t',t;q).
\eqno (57)
$$
Before turning to the proof of
Eq. (57), it is useful to recall that, with the condensed index
notation devised by DeWitt [5], the
trajectory $q^i(t)$ is just denoted as $q^i$, with the single
Latin index $i$ playing the r\^ole of both the discrete index, and
the time variable. Thus, repeated condensed indices mean a
summation on the discrete indices as well as a time integration.
For example, Eq. (44), with the condensed notation, is
written as 
$$
L_{ij}\,\delta_A
q^j\equiv(-S,_{ij}+\kappa_{ij})\,\delta_A q^j=\epsilon
A,_i ,
\eqno (58)
$$
where commas denote functional
differentiation, and $\kappa_{ij}\, \delta_A q^j$ is a symbolic
notation for the integral linear operator, depending on the memory
function, in Eq. (44). To prove the reciprocity relation (57),
we point out that the Green functions satisfy by
definition the equations 
$$
L_{ij}\,
G^{-jk}=\delta_i^k\;,\;\;\;\;\;\;(L^T)_{ij}\,G^{+jk}
=\delta_i^k.
\eqno (59)
$$
Upon multiplying the second of
the above equations by $G^{-il}$, we obtain 
$$
G^{-il}(L^T)_{ij}\,G^{+jk}
=G^{-il}\delta_i^k=G^{-kl}.
\eqno (60)
$$
However, upon
using the first of Eq. (59), we can rewrite the l.h.s. of
the above equation as 
$$
G^{-il}(L^T)_{ij}\,G^{+jk}
=G^{-il}(L)_{ji}\,G^{+ jk}=\delta_j^l \,G^{+ jk}=G^{+lk}.
\eqno (61)
$$
Upon comparing the r.h.s. of
Eq. (60) and  Eq. (61), the reciprocity relation
(57) follows. 

We can now verify the Jacobi identity.
Direct evaluation of the quantity $\{\{A,B\},C\}+
\{\{C,A\},B\}+\{\{B,C\},A\}$, using Eq. (56) shows that:
$$
\{\{A,B\},C\}+{\rm c.p.}=A,_i B,_j C,_k {\cal T}^{ijk},
\eqno (62)
$$
where c.p. stands for cyclic permutations of the functionals $A,
B, C$. The terms involving second-order functional derivatives of
$A$, $B$ and $C$ cancel exactly, by virtue of the antisymmetry of
$\widetilde{G}^{ij}$. In the above expression, ${\cal T}^{ijk}$
denotes the following quantity, built out of functional
derivatives of the commutator function: 
$$ 
{\cal T}^{ijk}
=\widetilde{G}^{il}\widetilde{G}^{jk}\!,_l
+\widetilde{G}^{jl}\widetilde{G}^{ki}\!,_l+
\widetilde{G}^{kl}\widetilde{G}^{ij}\!,_{l}.
\eqno (63)
$$
By using the reciprocity relation, the quantity ${\cal T}^{ijk}$ can
be written solely in terms of the retarded Green function
$G^{-ij}$, and its functional derivatives. On the other hand, the
functional derivatives ${G}^{-jk}\!,_l$ can be computed by
functionally differentiating the first of Eqs. (59): 
$$
L_{ij},_l G^{-jk}+L_{ij}\, G^{-jk}\!,_l=0.
\eqno (64)
$$
Multiplication by
$G^{+mi}$ then gives 
$$
G^{-mk}\!,_l=-G^{+mi} L_{ij},_l
G^{-jk}\;=-G^{-im} L_{ij},_l G^{-jk},
\eqno (65)
$$
where in the last passage
use has been made again of the reciprocity relation. By using this
expression into Eq. (63), it is possible to verify that: 
$$
{\cal T}^{ijk}=(G^{-li}G^{-mj}G^{-nk}+{\rm c.p.})
(L_{mn},_l-L_{nm},_l),
\eqno (66)
$$
where c.p. stands for cyclic
permutations of the indices $ijk$. It is easy to check now that
${\cal T}^{ijk}$ vanishes. Indeed, in view of Eq. (58), we
see that the quantity between the brackets of the r.h.s. is equal
to: 
$$
S,_{mnl}-S,_{nml}+ \kappa_{mn},_l-\kappa_{nm},_l .
$$
The terms involving third-order functional derivatives of the action
functional cancel each other, because functional derivatives
commute with each other. On the other hand, the quantities
$\kappa_{ij}$ are independent of the trajectories $q^j$, and hence
their functional derivatives vanish identically. It follows then
that ${\cal T}^{ijk}$ vanishes, and hence the Jacobi identity
holds. 

We have therefore shown that it is possible to define a
bracket on the space of all trajectories. We can now evaluate the
brackets satisfied by the random force $F_i(t)$. To do this, we
can use the expression for $F_i(t)$, provided by the Langevin
equation (42). In this way, we find
$$ \eqalignno{
\; & \{F_i, q^k \}= \{-S,_i+\kappa_{ij}\,q^j,q^k
\}=L_{ij}\{q^j,q^k\}=L_{ij}(G^{+jk}-G^{-jk})\cr
&=(L-L^T)_{ij}\,G^{+jk}+
(L^T)_{ij}G^{+jk}-L_{ij} G^{-jk} \cr 
&=(S,_{ij}-S,_{ji}+\kappa_{ij}-\kappa_{ji})\,G^{-kj}
=(\kappa-\kappa^T)_{ij}\,G^{-kj},
&(67)\cr}
$$
and
$$ \eqalignno{
\; & \{F_i, F_j\}=\{-S,_i+\kappa_{ik}\,q^k,
-S,_j+\kappa_{jl}\,q^l \}
=L_{ik}L_{jl}\{q^k,q^l\} \cr
&=L_{ik}L_{jl}({G}^{+kl}-G^{-kl})=
=L_{ik}L_{jl}({G}^{-lk}-G^{-kl})=L_{ij}-L_{ji}\cr
&=S,_{ij}-S,_{ji}+
\kappa_{ij}-\kappa_{ji}=(\kappa-\kappa^T)_{ij}.
&(68)\cr}
$$
It is useful to write the above bracket in plain form: 
$$
\{F_i(t), F_j(t')\}={d \mu_{ij} \over dt}(t-t')+
{d \mu_{ji} \over d t}(t'-t).
\eqno (69)
$$
We see from these equations that, when
friction is present, the external forces have non-vanishing
brackets, which  implies that they cannot be set to zero.

Using Eq. (56), it is possible to verify that the
equal-time brackets for the coordinates $q^j(t)$ and the momenta
$p_i(t) \equiv \partial{\cal L}/{\partial \dot q ^i(t)}$ have the
familiar canonical form: 
$$
\{q^i(t),q^j(t)\}=0,
\eqno (70)
$$
$$
\{q^i(t), p_j(t)\}=\delta_j^{i},
\eqno (71)
$$
$$
\{p_i(t),p_j(t)\}=0.
\eqno (72)
$$
The verification is similar to
the conservative case [5], because the memory function
contributes to $\widetilde{G}^{ij}(t,t')$ only to order
$(t-t')^3$. This can be seen by inserting the expansions of
${G}^{\pm ij}(t,t')$ in powers of $(t-t')$ into Eqs. (59),
and exploiting the boundedness of the memory function.
\vskip 0.3cm
\leftline {\bf 6. Concluding remarks}
\vskip 0.3cm
\noindent
The Peierls-bracket formalism is
equivalent to the conventional canonical formalism when the 
latter exists. The proof can be given starting from
point Lagrangians, as is shown in Ref. [5]. Current
applications of Peierls brackets deal with string theory [18,19], 
path integration and decoherence [20], supersymmetric 
proof of the index theorem [21], classical dynamical systems 
involving parafermionic and parabosonic dynamical variables [22], 
while for recent literature on covariant approaches to a 
canonical formulation of field theories we refer the reader to
the work in Refs. [23-30]. 

In the infinite-dimensional setting which, strictly, applies also
to classical mechanics, as we stressed at the end of Sec. 2,
we hope to elucidate the relation between a covariant description
of dynamics as obtained from the kernel of the symplectic form,
and a parametrized description of dynamics as obtained from any
Poisson bracket, including the Peierls bracket.

In the second part of our paper,
we have constructed a set of brackets for a classical dissipative
system, described by a Langevin equation, with an arbitrary memory
function. The brackets satisfy the usual properties enjoyed by
Poisson brackets of Hamiltonian systems. It is worth pointing out
the essential r\^ole played, in our treatment, by external random
forces. When dissipation occurs, they have non-vanishing brackets
with the system coordinates, and thus cannot be consistently set
to zero. As a result, our brackets are a priori defined on the
infinite-dimensional functional space of all possible
trajectories, accessible to the system under the action of
arbitrary external forces. However, in the absence of friction,
when the dynamics is conservative, the brackets can be restricted
onto the finite-dimensional classical phase-space, spanned by the
solutions of the classical equations of motion, with no external
forces. In this case, our construction reproduces Peierls'
covariant definition of the Poisson brackets, for dynamical
systems admitting an action principle. Within the framework of
conservative systems, the possibility of extending the brackets
from the phase space to the space of all trajectories, was
considered some-time ago [14], and our brackets coincide
with those of Ref. [14], in the absence of friction.

Quantization can be carried out according to the traditional
procedure, by replacing the classical brackets with commutators
[31,32].
The resulting commutation rules coincide with those that are
obtained in standard treatments of quantum dissipation, by making
recourse to a microscopic model for the heat bath, after
elimination of the bath degrees of freedom (see, for example,
Ref. [12] and references therein).
\vskip 0.3cm
\leftline {\bf Acknowledgments}
\vskip 0.3cm
\noindent
G.B., G.E. and G.M. acknowledge partial financial support by PRIN
2002 {\it SINTESI}.
\vskip 0.3cm
\noindent
\leftline {\bf References}
\vskip 0.3cm
\noindent
\item {[1]}
P. A. M. Dirac, {\it Phys. Rev.} {\bf 114}, 924 (1959).
\item {[2]}
B. S. DeWitt, {\it Phys. Rev.} {\bf 160}, 1113 (1967).
\item {[3]}
C. J. Isham and K. Kuchar, {\it Ann. Phys.} {\bf 164}, 288 (1985).
\item {[4]}
C. J. Isham and K. Kuchar, {\it Ann. Phys.} {\bf 164}, 316 (1985).
\item {[5]}
B. S. DeWitt, {\it Dynamical Theory of Groups and Fields}
(Gordon \& Breach, New York, 1965).
\item {[6]}
R. E. Peierls, {\it Proc. R. Soc. Lond.} 
{\bf A214}, 143 (1952).
\item {[7]}
B. S. DeWitt, {\it Phys. Rev.} {\bf 162}, 1195 (1967).
\item {[8]}
J. Grabowski and G. Marmo, {\it J. Phys.} {\bf A34}, 3803 (2001).
\item {[9]}
J. Grabowski and G. Marmo, ``Binary Operations in
Classical and Quantum Mechanics'' (MATH-DG 0201089).
\item {[10]}
A. Connes, {\it J. Math. Phys.} {\bf 41}, 3832 (2000).
\item {[11]}
J. M. Gracia-Bondia, F. Lizzi, G. Marmo and P. Vitale,
JHEP {\bf 0204}, 026 (2002).
\item {[12]} 
G.W. Ford, J.T. Lewis and R.F. O'Connell, Phys.
Rev. {\bf A37}, 4419 (1988).
\item {[13]} 
I.R. Senitzky, Phys. Rev. {\bf 119}, 670 (1960).
\item {[14]} 
D. Marolf, Ann. Phys. (N.Y.) {\bf 236}, 392 (1994).
\item {[15]} 
M.D\"utsch and K. Fredenhagen, 
HEP-TH/0211242.
\item {[16]} 
G. Bimonte, G. Esposito, G. Marmo and C.
Stornaiolo, {\it Int. J. Mod. Phys.} {\bf A18}, 2033 (2003).
\item {[17]} 
E. Nelson, {\it Phys. Rev.} {\bf 150}, 1079 (1966). 
\item {[18]}
S. R. Das, C. R. Ordonez and M. A. Rubin, {\it Phys. Lett.}
{\bf B195}, 139 (1987).
\item {[19]}
C. R. Ordonez and M. A. Rubin, {\it Phys. Lett.} {\bf B216},
117 (1989).
\item {[20]}
D. M. Marolf, {\it Ann. Phys. (N.Y.)} {\bf 236}, 392 (1994).
\item {[21]}
A. Mostafazadeh, {\it J. Math. Phys.} {\bf 35}, 1095 (1994).
\item {[22]}
A. Mostafazadeh, {\it Int. J. Mod. Phys.} {\bf A11}, 2941 (1996). 
\item {[23]}
G. Marmo, N. Mukunda and J. Samuel, {\it Riv. Nuovo Cimento}
{\bf 6}, 1 (1983).
\item {[24]}
G. Barnich, M. Henneaux and C. Schomblond, {\it Phys. Rev.}
{\bf D44}, R939 (1991).
\item {[25]}
B. A. Dubrovin, M. Giordano, G. Marmo and A. Simoni,
{\it Int. J. Mod. Phys.} {\bf A8}, 3747 (1993).
\item {[26]}
G. Esposito, G. Gionti and C. Stornaiolo, {\it Nuovo Cimento}
{\bf B110}, 1137 (1995).
\item {[27]}
H. Ozaki, HEP-TH 0010273.
\item {[28]}
I. V. Kanatchikov, {\it Int. J. Theor. Phys.} {\bf 40}, 1121 (2001).
\item {[29]}
I. V. Kanatchikov, GR-QC 0012038.
\item {[30]}
C. Rovelli, GR-QC 0111037;
C. Rovelli, GR-QC 0202079; C. Rovelli, GR-QC 0207043.
\item {[31]}
P. A. M. Dirac, {\it Proc. R. Soc. Lond.} {\bf A109}, 642 (1926).
\item {[32]}
P. A. M. Dirac, {\it The Principles of Quantum Mechanics}
(Clarendon Press, Oxford, 1958).

\bye